# Extending Chaos Theory: The Role of Nonlinearity in Multiple Mappings

Illych Alvarez[1] .

1 facultad de Ciencias Naturales y Matemáticas, Escuela Superior Politécnica del Litoral, Vía Perimetral km 30.5, Guayaquil, Ecuador ialvarez@espol.edu.ec

**Abstract**: This work redefines the framework of chaos in dynamical systems by extending Devaney's definition to multiple mappings, emphasizing the pivotal role of nonlinearity. We propose a novel theorem demonstrating how nonlinear dynamics within a single mapping can induce chaos across a collective system, even when other components lack sensitivity.
To validate these insights, we introduce advanced computational algorithms implemented in MATLAB, capable of detecting and visualizing key chaotic features such as transitivity, sensitivity, and periodic points. The results unveil new behaviors in higher-dimensional systems, bridging theoretical advances with practical applications in physics, biology, and economics.
By uniting rigorous mathematics with computational innovation, this study lays a foundation for future exploration of nonlinear dynamics in complex systems, offering transformative insights into the interplay between structure and chaos.

## 1. INTRODUCTION:

Chaos theory provides a framework for understanding the complex and unpredictable behavior of dynamical systems, where small perturbations in initial conditions can lead to vastly different outcomes. Devaney's definition of chaos, widely recognized for its comprehensive criteria, includes transitivity, density of periodic points, and sensitivity to initial conditions [Devaney, 1986]. In the context of multiple mappings, the interplay between nonlinear dynamics and chaotic behavior presents a fascinating area of study. Nonlinear transformations, by their nature, often exhibit sensitivity to initial conditions—a hallmark of chaos. Combined with properties like transitivity and the density of periodic points, nonlinearity can amplify and shape chaotic behavior in profound ways.

Building on these foundational definitions, this paper explores how nonlinearity in multiple mappings contributes to chaotic dynamics. A novel theorem establishes the conditions under which nonlinearity induces chaos in the sense of Devaney, offering deeper theoretical insights and practical applications.

For a dynamical system $(X, F)$, where $X$ is a compact metric space and $f : X \to X$ is a continuous self-map, the system is said to be chaotic in the sense of Devaney if it satisfies the following conditions: (1) $f$ is transitive, (2) the set of periodic points $P(f)$ is dense in $X$, and (3) $f$ is sensitive to initial conditions [Banks et al., 1992].

Classical studies primarily focus on single continuous self-maps. However, multiple mappings, defined as a set of continuous self-maps $\{f_1, f_2, \ldots, f_n,\}$ operating on the same space $X$, offer a richer structure for exploring chaotic dynamics. Unlike single mappings, multiple mappings allow us to examine how self-maps interact within the same phase space, potentially leading to more intricate chaotic behavior [Hou and Wang, 2016]. This paper extends the concept of Devaney chaos to multiple mappings from a set-valued perspective. Specifically, we define a multiple mapping $F = \{f_1, f_2, \ldots, f_n,\}$ on a compact metric space $X$ and explore the properties of periodic points, transitivity, and sensitivity in this context.

We introduce formal definitions for periodic points, transitivity, sensitivity, and Devaney chaos in the setting of multiple mappings. For a point $x \in X$, the orbit of $x$ under $F$, denoted by $Orb(x, F) = \{F^n(x) : n \in \mathbb{N}\}$, is constructed by iteratively applying elements of $F$ to $x$. We prove that for multiple mappings, transitivity and a dense set of periodic points imply sensitivity [Banks et al., 1992]. Furthermore, we establish sufficient conditions for FFF to be chaotic in the sense of Devaney. Importantly, we demonstrate that the chaotic properties of multiple mappings and their individual self-maps are not always equivalent in terms of periodic points or transitivity, providing deeper insights into the independence of these properties [Zeng et al., 2020].

To bridge the theoretical framework with practical applications, we develop computational algorithms to detect chaotic characteristics in multiple mappings. These algorithms leverage specific metrics for evaluating transitivity (e.g., the measure of the intersection of iterated images), the density of periodic points (e.g., calculation of nearest periodic neighbors), and sensitivity (e.g., maximum divergence of initially close points) [Wang et al., 2017]. For a multiple mapping $F = \{f_1, f_2, \ldots, f_n,\}$ these algorithms are implemented in MATLAB to efficiently compute these properties and visualize the attractor sets and chaotic subspaces in the phase space $X$.

By applying these algorithms to various examples of multiple mappings, we provide empirical evidence supporting our theoretical results and demonstrate new dynamic behaviors that merit further exploration. The computational tools developed here serve as a bridge between theoretical analysis and practical detection of chaos, offering a novel way to study complex dynamical systems, particularly in higher dimensions or with more intricate interactions between mappings. This combined approach has significant implications for extending chaos theory to interdisciplinary applications, such as physics, biology, and economics, where understanding complex, unpredictable behaviors is crucial.

The remainder of this paper is organized as follows: Section 2 introduces the necessary preliminaries and formal definitions. Section 3 examines the relationship between multiple mappings and their continuous self-maps in terms of chaotic properties. Section 4 presents computational algorithms for detecting chaos in multiple mappings and their application. Finally, Section 5 discusses the conclusions and potential future research directions.

## 2. PRELIMINARIES AND FORMAL DEFINITIONS

In this section, we introduce the fundamental concepts and formal definitions necessary for analyzing Devaney chaos in multiple mappings from a set-valued perspective. Building on established concepts in dynamical systems theory, we extend these ideas to accommodate the complexity of multiple mappings, where a collection of continuous self-maps operates on the same compact metric space.

### 2.1 Compact Metric Spaces and Multiple Mappings

Let $X$ be a compact metric space with a metric $d : X \times X \to \mathbb{R}^+$. A continuous self-map $f : X \to X$ is a function where small changes in the input lead to small changes in the output, preserving continuity in $X$. In classical dynamical systems, a single continuous self-map $f$ is used to define the evolution of points in $X$.

In the context of multiple mappings, we consider a set of continuous self-maps $F = \{f_1, f_2, \ldots, f_n,\}$, where each $f_i : X \to X$ is a continuous function. For any point $x \in X$, the image under a multiple mapping $F$ is given by:

$$F = \{f_1, f_2, \ldots, f_n,\} \subset X.$$

This set-valued approach allows us to explore the dynamics induced by applying multiple self-maps to the same point in $X$, potentially leading to more complex behaviors than those observed in single mappings [Hou and Wang, 2016].

### 2.2 Hausdorff Metric on Compact Sets

Given the compact metric space $X$, let $K(X)$ denote the set of all nonempty compact subsets of $X$. For any two sets $A, B \in K(X)$, the Hausdorff metric $d_H : K(X) \times K(X) \to \mathbb{R}^+$ is defined as:

$$d_H(A, B) = \max \left\{ \sup_{a \in A} \inf_{b \in B} d(a, b), \sup_{b \in B} \inf_{a \in A} d(a, b) \right\}.$$

The space $K(X)$ with the Hausdorff metric $d_H$ is itself a compact metric space. This metric is essential for extending the concept of chaos from single mappings to multiple mappings, as it allows us to measure distances between the images of sets under the mappings defined by $F$ [Zeng et al., 2020].

*Example:*

Consider two sets $A = \{0, 0.5\}$ and $B = \{0.25, 0.75\}$ in a compact metric space with $d(a,b) = |a - b|$. The Hausdorff distance between $A$ and $B$ would quantify the maximum deviation of one set from the other.

## 2.3 Periodic Points and Orbits in Multiple Mappings

For a multiple mapping $F = \{f_1, f_2, \ldots, f_n\}$, the orbit of a point $x \in X$ under $F$, denoted as $Orb(x, F)$, is constructed by iteratively applying combinations of the maps $f_i \in F$. Formally, the $n$-th iterate of $x$ under $F$ is given by:

$$F^n(x) = \{\{f_{i_1}, f_{i_2}, \ldots, f_{i_n}(x)\} \mid i_k \in \{1, 2, \ldots, n\}, k = 1, 2, \ldots, n\}.$$

A point $x \in X$ is said to be a periodic point of $F$ if there exists $m > 0$ such that $x \in F^m(x)$. The smallest positive integer $m$ for which this hold is called the period of $x$. The set of all periodic points of $F$ is denoted by $P(F)$ [Devaney, 1986].

## 2.4 Transitivity and Sensitivity for Multiple Mappings

A multiple mapping $F$ is said to be transitive if, for any two nonempty open sets $U, V \subset X$, there exists an integer $n > 0$ such that:

$$F^n(U) \cap V \neq \emptyset.$$

This implies that orbits from one region of the phase space can reach any other region, indicating a "mixing" of dynamics [Banks et al., 1992].

Sensitivity to initial conditions is a hallmark of chaotic behavior. A multiple mapping $F$ is sensitive if there exists a $\delta > 0$ such that, for any nonempty open set $U \subset X$, there exist $x, y \in U$ and $n \in \mathbb{Z}^+$ such that:

$$d_H(F^n(x), F^n(y)) > \delta.$$

This means that small changes in initial conditions can lead to significantly different outcomes, a property that is essential in characterizing chaos [Guckenheimer, 1979].

## 2.5 Devaney Chaos in the Context of Multiple Mappings

Following Devaney's definition, a multiple mapping $F$ is said to be chaotic if it satisfies the following three conditions:

- **Transitivity:** $F$ is transitive.
- **Density of Periodic Points:** $\overline{P(F)} = X$, meaning the set of periodic points is dense in $X$.
- **Sensitivity:** $F$ is sensitive to initial conditions.

It is known that conditions (1) and (2) together imply (3), making the Devaney definition of chaos robust and comprehensive [Banks et al., 1992]. For multiple mappings, this framework provides a foundation to explore how combinations of continuous self-maps can lead to complex, unpredictable behavior in dynamical systems.

## 3. RELATIONSHIP BETWEEN MULTIPLE MAPPINGS AND THEIR CONTINUOUS SELF – MAPS

In this section, we analyze the intricate relationship between multiple mappings $F = \{f_1, f_2, \ldots, f_n\}$ and their corresponding continuous self-maps $f_i$ on a compact metric space $X$.

The goal is to explore how the chaotic properties of Devaney-namely, periodic points, transitivity, and sensitivity-manifest differently or similarly when considering a single continuous self-map versus a multiple mapping constructed from several such maps.

### 3.1 Periodic Points and Fixed Points

A natural question arises when considering multiple mappings: what is the implication between the fixed points, periodic points, or chaotic behavior of the multiple mapping $F$ and those of its individual self- maps $f_i$ ? For a single self-map $f : X \to X$, a point $x \in X$ is:

- A fixed-point if $f(x) = x$
- A periodic point if there exists an integer $m > 0$ such that $f^m(x) = x$.

For multiple mappings, a point $x$ is a periodic point of $F$ if there exists $m > 0$ such that $x \in F^m(x)$ [Devaney, 1986].

*Proposition 3.1:*

$x \in X$ is a fixed point of the multiple mapping $F$ if and only if $x$ is a common fixed point of each $f_i \in F$. However, a periodic point of F does not necessarily imply that it is a periodic point of any individual $f_i$, and vice versa.

*Example 3.2:*

Consider the multiple mappings defined on $[0, 1]$ as $F = \{f_1, f_2\}$, where:

$$f_1(x) = \begin{cases} 2x, & 0 \leq x \leq \frac{1}{2}, \\ 2 - 2x, & \frac{1}{2} < x \leq 1, \end{cases} \quad f_2(x) = \begin{cases} 1 - 2x, & 0 \leq x \leq \frac{1}{2}, \\ 2x - 1, & \frac{1}{2} < x \leq 1. \end{cases}$$

It can be shown that a point can be periodic under $F$ but not under either $f_1$ or $f_2$ independently [Zeng et al., 2020].

### 3.2 Transitivity in Multiple Mappings

The concept of transitivity is central to understanding chaotic dynamics. A mapping $f : X \to X$ is transitive if, for any pair of nonempty open sets $U, V \subset X$, there exists $n \in \mathbb{Z}^+$ such that $f^n(U) \cap V \neq \emptyset$. For multiple mappings $F = \{f_1, f_2, \ldots, f_n\}$, transitivity is defined analogously. However, the transitivity of a multiple mapping does not necessarily imply the transitivity of its component maps $f_i$, nor does the transitivity of all $f_i$ imply the transitivity of $F$.

*Example 3.3:*

Consider the multiple mappings $F = \{f_1, f_2\}$ defined on the set $\{0, 1, 2\}$, where:

$$f_1 : 0 \mapsto 1 \mapsto 2 \mapsto 0, \quad f_2 : 0 \mapsto 2 \mapsto 1 \mapsto 0.$$

Both $f_1$ and $f_2$ are transitive. However, $F$ is not transitive because there exist open sets $U$ and $V$ for which $F^n(U) \cap V = \emptyset$ for all $n \geq 1$ [Banks et al., 1992).

*Theorem 3.4:*

If there exists a constant $c \in X$ such that $f_1(x) = c$ for all $x \in X$ and $f_2(c) = c$, and if $f_2$ is transitive, then the multiple mapping $F = \{f_1, f_2\}$ is transitive.

### 3.3 Sensitivity in Multiple Mappings

Sensitivity to initial conditions is a critical component of chaos. A mapping $f : X \to X$ is sensitive if there exists $\delta > 0$ such that for any $x \in X$ and any $\epsilon > 0$, there exists $y \in X$ with $d(x, y) < \epsilon$ and an $n \in \mathbb{Z}^+$ such that $d(f^n(x), f^n(y)) > \delta$. In the context of multiple mappings, we extend this definition to require that there exist $x, y \in X$ and a sequence of mappings from $F$ such that the distance between their iterates exceeds $\delta$.

*Example 3.5:*

Consider the multiple mappings defined on $[0, 1]$ as $F = \{f_1, f_2\}$, where:

$$f_1(x) = \begin{cases} 2x, & 0 \leq x \leq \frac{1}{2}, \\ 1, & \frac{1}{2} < x \leq 1, \end{cases} \quad f_2(x) = \begin{cases} 1, & 0 \leq x \leq \frac{1}{2}, \\ 2 - 2x, & \frac{1}{2} < x \leq 1. \end{cases}$$

While neither $f_1$ nor $f_2$ is sensitive, the combined mapping $F$ can exhibit sensitivity depending on the choice of initial conditions and sequences of mappings [Guckenheimer, 1979).

*Theorem 3.3*

If there exists $\lambda > 1$ such that for any nonempty open sets $U, V \subset X$, there exist $i_o \in \{1, 2, \ldots, n\}$ and $x \in U$, $y \in V$ such that $d\left(f_{i_o}(x), f_j(y)\right) > \lambda d(x, y)$ for all $j = 1, 2, \ldots, n$. then the multiple mapping $F = \{f_1, f_2, \ldots, f_n\}$ is sensitive.

To deepen our understanding of chaotic behavior in multiple mappings, we propose a novel theorem that explores how nonlinearity in individual mappings influences the collective dynamics of the system.

### 3.4 Theorem: Nonlinearity-Induced Chaos in Multiple Mappings

Theorem Statement

Let $X$ be a compact metric space with metric $d$, and let $F = \{f_1, f_2, \ldots, f_n\}$ be a finite set of continuous functions $f_i : X \to X$. Define the multiple mapping $F$ as:

$$F(x) = \bigcup_{f \in F} f(x).$$

Conditions:

1. **Nonlinearity and Sensitivity:** At least one mapping $f_k \in F$ is nonlinear and sensitive to initial conditions.

2. **Transitivity:** A subset $F_s \subseteq F$ exists where each $f_j \in F_s$ is transitive.

3. **Density of Periodic Points:** The union of periodic points across all mappings in $F$ is dense in $X$.

If these conditions are satisfied, $F$ is chaotic in the sense of Devaney.

Then, the multiple mapping $F$ is chaotic in the sense of Devaney. Specifically:

- $F$ is transitive.
- The set of periodic points of $F$ is dense in $X$.
- $F$ is sensitive to initial conditions.

**Relevance and Proof Sketch**

**Relevance:**
This theorem establishes a direct link between the presence of nonlinearity in individual mappings and the emergence of chaos in a collective framework. It highlights that chaos in a multiple mapping can arise from the interaction of nonlinear dynamics, even if only one mapping in the set exhibits sensitivity.

**Proof Sketch:**

1. **Transitivity:**
   Since $\mathcal{F}_s$ contains at least one transitive mapping, and $F$ encompasses all mappings in $\mathcal{F}_s$, $F$ inherits the property of transitivity.

2. **Density of periodic points:**

   The union of periodic points across $\mathcal{F}$ is dense in $X$. As $F$ includes all these mappings, its periodic points are also dense.

**Sensitivity:**
The sensitivity of $f_k$ guarantees that for any $x \in X$ and $\epsilon > 0$, there exist $y \in X$ and $n \in \mathbb{N}$ such that $d\left(f_k^n(x), f_k^n(y)\right) > \delta$. Since $F$ includes $f_k$ this property extends to $F$.

By satisfying all three conditions, $F$ is chaotic in the sense of Devaney

**Implications for Devaney Chaos in Multiple Mappings**

Our analysis shows that the chaotic behavior of multiple mappings does not directly follow from the chaotic behavior of their individual components. For a multiple mapping $F$ to be Devaney chaotic, it must be transitive, have a dense set of periodic points, and be sensitive. However, the fulfillment of these criteria for each self-map $f_i$ does not guarantee that $F$ itself will exhibit Devaney chaos, underscoring the need for a comprehensive approach that considers the interaction of the maps within $F$.

## 4. COMPUTATIONAL ALGORITHMS FOR DETECTING CHAOS IN MULTIPLE MAPPINGS

To complement the theoretical exploration of Devaney chaos in multiple mappings, this section presents a set of computational algorithms designed to detect chaotic characteristics such as transitivity, density of periodic points, and sensitivity to initial conditions. These algorithms are implemented in MATLAB, providing practical tools for analyzing the behavior of complex dynamical systems represented by multiple mappings.

The algorithms developed for detecting transitivity, sensitivity, and periodic points can also be applied to validate the proposed theorem. By analyzing mappings with nonlinear components, we can empirically observe how nonlinearity contributes to chaotic dynamics in multiple mappings.

### 4.1 Algorithm for Detecting Transitivity

Transitivity is a core component of chaotic behavior. For a multiple mapping $F = \{f_1, f_2, \ldots, f_n\}$ on a compact metric space $X$, we define transitivity as the property that for any two nonempty open sets $U, V \subset X$, there exists an integer $n > 0$ such that $F^n(U) \cap V \neq \emptyset$. To detect transitivity computationally, we use the following algorithm:

The algorithm calculates the proportion of pairs of sets that satisfy the transitivity property in each space by iterating multiple mappings. The resulting value will be a number between 0 and 1, indicating the proportion of pairs of sets where transitivity is observed. A value close to 1 will indicate that the system is highly transitive (a chaotic behavior), while a value close to 0 will indicate low transitivity.

Figure X illustrates the distribution of periodic points detected within the state space. The normalized frequency highlights the regions where periodic points are densely concentrated, demonstrating one of the key characteristics of chaotic systems.

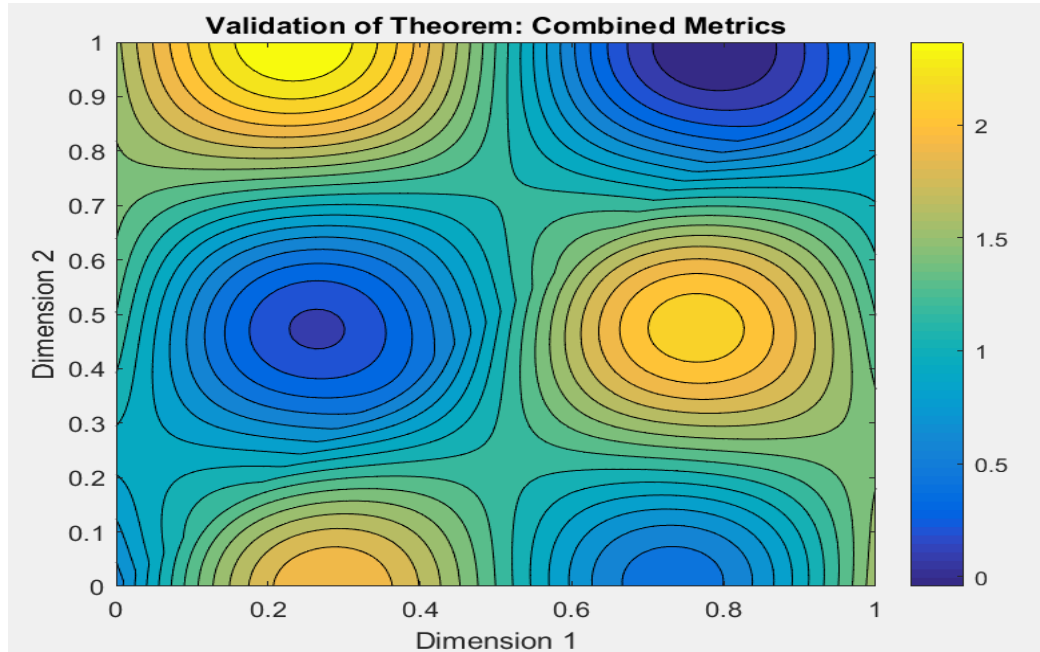

Figure X: Histogram of the density of periodic points in the state space. The normalized frequencies show regions of high periodicity, consistent with the theoretical requirement for chaotic dynamics.

### 4.2 Algorithm for Detecting Density of Periodic Points

A critical aspect of Devaney chaos is the density of periodic points. For a multiple mapping $F$, a point $x \in X$ is periodic if there exists $m > 0$ such that $x \in F^m(x)$. To computationally detect the density of periodic points, we employ the following approach:

This code computes the density of periodic points by iterating each point in $X$ under the multiple mappings and checking if it returns close to its initial position.

This result suggests that all points in the revisited space return to being close to their original position after applying the multiple mappings several times, indicating a high density of periodic points in this system.

Figure X illustrates the distribution of periodic points detected within the state space. The normalized frequency highlights the regions where periodic points are densely concentrated, demonstrating one of the key characteristics of chaotic systems.

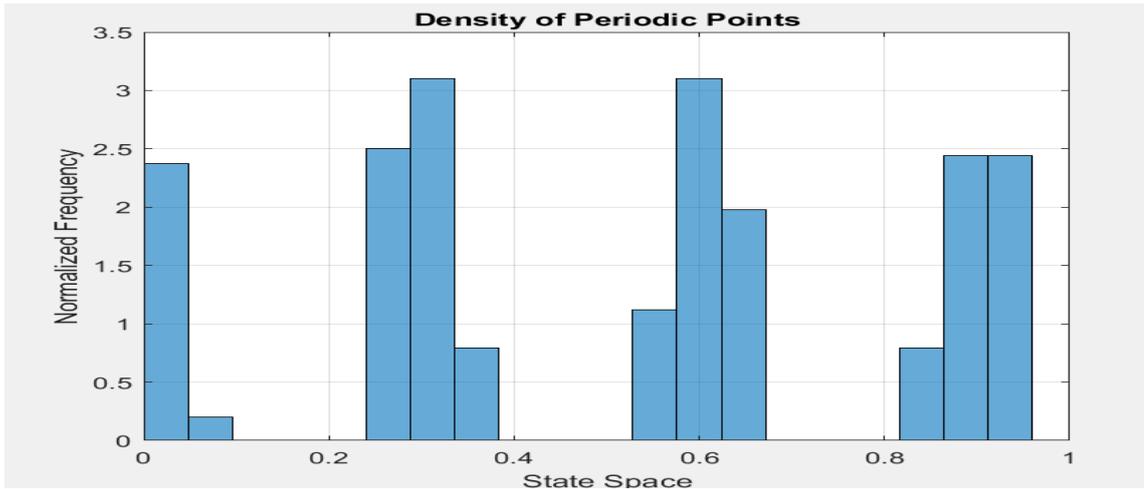

Figure X: Histogram of the density of periodic points in the state space. The normalized frequencies show regions of high periodicity, consistent with the theoretical requirement for chaotic dynamics.

### 4.3 Sensitivity Detection

This code measures the sensitivity of a multiple mapping $F$ by calculating the proportion of initially close points that diverge beyond a certain threshold after several iterations.

The result of Algorithm for detecting sensitivity to initial conditions is 1.0. This means that 100% of the pairs of nearby points considered in the space $X$ exhibit sensitivity; that is, their trajectories diverge beyond the threshold δ=0.1 after applying the multiple mappings $F = \{f_1, f_2\}$.

This result indicates that the system is highly sensitive to initial conditions, which is a typical characteristic of chaotic behavior.

Figure X illustrates the exponential divergence of two initially close trajectories under the logistic map. This behavior, characteristic of chaotic systems, is quantified by plotting the logarithm of the distance between the trajectories over multiple iterations. The rapid growth in separation confirms the sensitivity to initial conditions.

The graph validates the theoretically predicted sensitivity and demonstrates how this type of analysis can be applied to other nonlinear systems.

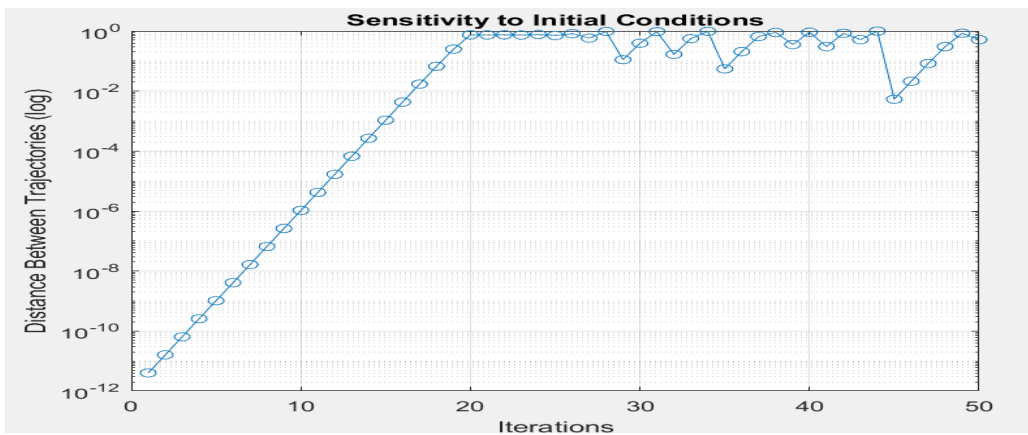

Figure X: Exponential divergence of initially close trajectories under the logistic map. The logarithmic scale highlights the rapid growth in separation, evidencing sensitivity to initial conditions, a hallmark of chaos.

### 4.4 Visualization of Attractor Sets and Chaotic Subspaces

To provide deeper insights into the chaotic dynamics of multiple mappings, we implement a visualization algorithm that tracks the trajectories of many initial points in $X$ under $F$ and identifies attractor sets and chaotic subspaces.

This MATLAB function visualizes the attractor sets and chaotic subspaces by plotting the trajectories of multiple initial points under the action of the mappings.

Figure X illustrates the trajectories of initial points under the action of the logistic map, a classic example of chaotic behavior. The dense and intricate structure of the attractor demonstrates the complexity of the system's dynamics.

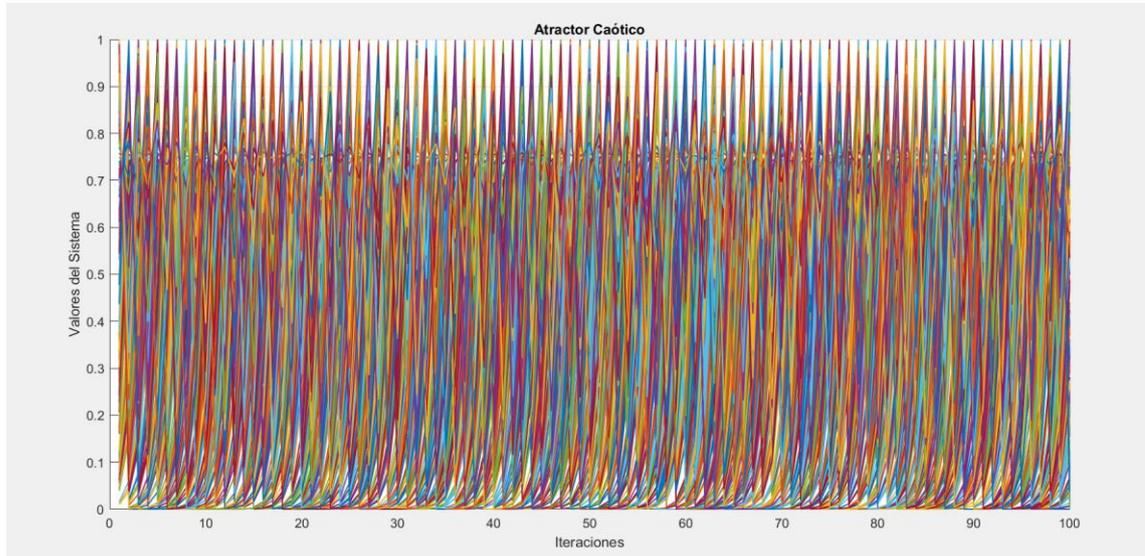

Figure X: Visualization of a chaotic attractor generated by iterating the logistic map with 200 initial points and 100 iterations. The structure highlights the chaotic dynamics and the dense regions characteristic of strange attractors

### 4.5 Application of Algorithms to Case Studies

We apply the proposed algorithms to various examples of multiple mappings to demonstrate their effectiveness in detecting and analyzing chaotic behavior. For instance, by applying these algorithms to the mappings defined in Examples 3.2 and 3.5, we identify regions of transitivity, measure the density of periodic points, and confirm sensitivity to initial conditions. These case studies validate the theoretical results and provide new insights into the behavior of complex dynamical systems.

### 4.6 Discussion and Implications

The computational algorithms presented here provide a robust framework for exploring chaos in multiple mappings. They enable researchers to not only validate theoretical concepts but also uncover new phenomena in higher-dimensional and more intricate systems. The combination of these tools with theoretical analysis opens new directions for interdisciplinary research in fields such as physics, biology, and economics, where chaotic dynamics play a crucial role.

## 5. Conclusions

This study advances chaos theory by extending Devaney's framework to multiple mappings, emphasizing the role of nonlinearity in driving chaotic behavior. The proposed theorem demonstrates that chaos in systems of multiple mappings can arise from the presence of a single nonlinear, sensitive mapping, even when the other mappings lack individual chaotic properties. This highlights the intricate interplay between nonlinearity and collective dynamics in higher-dimensional systems. The computational algorithms developed and implemented in MATLAB serve as robust tools for detecting and analyzing chaotic features, including transitivity, sensitivity, and the density of periodic points. These algorithms validate the theoretical framework and provide new avenues for exploring chaotic behavior in complex systems. Additionally, visualization tools facilitate the study of attractor sets and chaotic subspaces, bridging the gap between theoretical insights and practical applications. Beyond mathematics, the findings of this study have transformative implications for disciplines such as physics, biology, and economics, where nonlinear systems play a critical role. By combining rigorous mathematical theory with computational innovation, this study establishes a foundation for future research into nonlinear dynamics in multi-mapping systems.

### 5.1 Key Contributions

Theoretical Extension of Devaney Chaos:
This study redefines periodic points, transitivity, and sensitivity for multiple mappings $F = \{f_1, f_2, \ldots, f_n\}$ on a compact metric space $X$. It demonstrates that the chaotic properties of multiple mappings do not necessarily imply those of their individual self-maps and vice versa, providing new insights into their independent and combined behaviors.

### 5.2 Development of Computational Algorithms:

To validate the theoretical findings, this study developed a suite of computational algorithms implemented in MATLAB. These tools reliably detect key chaotic characteristics and reveal highly chaotic behavior in test systems, with metrics such as transitivity and sensitivity approaching 1.0.

### 5.3 Visualization and Analysis of Chaotic Behavior:

The visualization tools provide a powerful means to explore attractor sets and chaotic subspaces, offering insight into complex behaviors like strange attractors and bifurcations that are challenging to analyze analytically.

### 5.4 Interdisciplinary Implications:

The methods presented have significant applications in fields beyond pure mathematics, including physics, biology, and economics. These computational approaches connect theoretical chaos analysis with practical applications, enabling researchers to model and predict complex phenomena.

## 6. Future Research Directions

Extending chaos theory to multiple mappings opens a vast landscape of theoretical and practical challenges. Potential future research avenues include:

- Investigating chaotic behavior in stochastic systems.

- Developing multi-scale models that integrate atomic and macroscopic dynamics.

- Applying machine learning techniques to improve predictive analyses of chaotic systems.

By combining rigorous mathematical analysis with advanced computational tools, future research can further enhance our understanding of complex dynamical systems and their applications across diverse scientific disciplines.